\documentclass[12pt,english,floatfix,showkeys,superscriptaddress,aps,prd,preprint]{revtex4}

\usepackage[english]{babel}
\usepackage[T1]{fontenc}
\usepackage[utf8]{inputenc}
\usepackage{amsmath}
\usepackage{amssymb}
\usepackage{amsbsy}
\usepackage{amsfonts}
\usepackage{amsopn}
\usepackage{amstext} 
\usepackage{amssymb}
\usepackage{amsfonts}  
\usepackage{color}
\usepackage{slashed}
\usepackage{esint}
\usepackage{graphicx} 
\usepackage{float}
\usepackage{units}
\usepackage{textcomp}
\usepackage[shortlabels]{enumitem}
\usepackage{hyperref}
\usepackage{tensor}
\usepackage{tcolorbox}
\usepackage{booktabs} 
\usepackage{caption,subcaption} 
 \usepackage{xcolor}

\begin{document}

\title{Dymnikova Black Hole Surrounded by Quintessence }

\author{M. H. Mac\^{e}do}
\email{matheus.macedo@fisica.ufc.br}
\affiliation{Universidade Federal do Cear\'a (UFC), Departamento de F\'isica,\\ Campus do Pici, Fortaleza - CE, C.P. 6030, 60455-760 - Brazil.}
%%%%%%%%%%%%%%%%%%%%%%%%%%%%%%%%%%%%%%%%%%%%%%%%%%%%%%%%%%%%%%%%%%%%%%
\author{J. Furtado}
\email{job.furtado@ufca.edu.br}
\affiliation{Universidade Federal do Cariri (UFCA), Av. Tenente Raimundo Rocha, \\ Cidade Universit\'{a}ria, Juazeiro do Norte, Cear\'{a}, CEP 63048-080, Brasil} 
%%%%%%%%%%%%%%%%%%%%%%%%%%%%%%%%%%%%%%%%%%%%%%%%%%%%%%%%%%%%%%%%%%%%%%%
 
\author{R. R. Landim}
\email{renan@fisica.ufc.br}
\affiliation{Universidade Federal do Cear\'a (UFC), Departamento de F\'isica,\\ Campus do Pici, Fortaleza - CE, C.P. 6030, 60455-760 - Brazil.}

\begin{abstract}

The Dymnikova black hole (BH) is a regular solution that interpolates between a de Sitter core near the origin and a Schwarzschild-like behavior at large distances. In this work, we investigate the properties of a Dymnikova BH immersed in a quintessential field, characterized by the state parameter $\omega$ and a normalization constant $c$. We explore the thermodynamic behavior, null geodesics, scalar quasinormal modes and shadow profiles for this model. Our analysis shows that the presence of quintessence alters the Hawking temperature and specific heat, leading to parameter-dependent phase transitions. The null geodesics and corresponding black hole shadows are also found to be sensitive to the model parameters, especially $\omega$ and $c$. This sensitivity influences light deflection and shadow size. Furthermore, we compute the scalar quasinormal modes and observe that quintessence tends to enhance the damping of the modes, indicating greater stability under perturbations. 

\end{abstract}


\maketitle
%%%%%%%%%%%%%%%%%%%%%%%%%%%%%%%%%%%%%%%%%%%%%%%%%%%%%%%%%%%%%%%%%%%%%%%%%%%%%%%
\section{Introduction}

The first solution of Einstein equations was discovered by Schwarzshild in 1916, such solution describes a space-time spherically symmetric, with non-rotating and chargeless which is generated by an object known as black hole.  Black holes have become one of the most important solutions in General Relativity (GR). Until then, the theory had been well tested in weak-field conditions, such as solar system \cite{Will:2001mx}, and some moderately relativistic scenarios, as produced by a white dwarf pulsar \cite{Pelisoli_2022}. However, the detection of gravitational waves by LIGO and the first images of black holes have tested the predictions of GR in extreme environments \cite{LIGOScientific:2016aoc,EventHorizonTelescope:2019dse}.

Sakharov and Gliner \cite{Sakharov:1966,Gliner:1966} proposed that essential singularities, such as the one present at the origin of the Schwarzschild black hole, could be avoided if the standard vacuum with a zero energy-momentum tensor were replaced by a vacuum-like medium possessing a de Sitter metric. After that, several singularity-free solutions, called regular solutions, were proposed. Bardeen replaced the black hole's constant mass with a position-dependent mass \cite{bardeen:1968}, while other solutions, such as the Simpson-Visser solution, have a structure that can represent both a regular black hole and a wormhole \cite{Simpson:2018tsi} depending on the choice of the parameters. Another important regular solution was proposed by Dymnikova, based on the gravitational analogue of the Schwinger effect \cite{Dymnikova:1992ux}.

The gravitational analogue of the Schwinger effect, proposed by Dymnikova \cite{Dymnikova:1992ux}, represents a vacuum solution that, at large distances from the origin, behaves like the Schwarzschild solution, while near the core, it reduces to the De Sitter solution. Recently, the Dymnikova black hole has been studied in various contexts, e.g., in the context of pure lovelock gravity \cite{Estrada:2024uuu}, the generalized uncertainty principle \cite{Ma:2024tqp}, theories with higher-curvature corrections \cite{Konoplya:2024kih} and wormhole solutions \cite{Estrada:2023pny}. Among the topics investigated within the context of Dymnikova black hole are quasinormal modes and thermodynamic properties in higher dimensions discussed in \cite{Macedo:2024dqb}. 

On one hand, the thermodynamic properties allows the exploration of black hole evaporation, which arises through the creation and annihilation of particles near the event horizon. In this process, one particle may fall toward the center while the other escapes outward. This phenomenon, known as Hawking radiation, is responsible for the gradual mass loss of the black hole over time \cite{Hawking:1975vcx}. Thus, although at first glance it seemed meaningless to associate temperature with an object from which nothing can escape \cite{Wald:1999vt}, Hawking and Bekenstein demonstrated that black holes obey laws analogus to those of classical thermodynamics \cite{Bekenstein:1972tm,Bekenstein:1973ur,Bekenstein:1974ax,Bardeen:1973gs}, with quantities similar to temperature, entropy, heat capacity, and others. Over the years, extensive research have been conducted for different black holes, particularly for regular black holes such as those of Hayward, Bardeen, Dymnikova, and Simpson-Visser \cite{Molina:2021hgx,Ali:2018boy,Dymnikova:1996,Nosirov:2023ism}.

On the other hand, Regge amd Wheeler \cite{Regge:1957td}, while studying the Schwarzschild black hole stability, showed that metric perturbation lead to second-order differential equations, which take a form similar to the Schrödinger equation, with the difference of the presence of a damping term. Such oscillations, called quasinormal modes, are characterized by complex frequencies and describe the gravitational wave signals emitted by black holes after merging with another black hole or an interaction with gravitational waves \cite{Berti:2009kk}. The real component of these frequencies determines the oscillation rate, whereas the imaginary part governs the decay of the signal over time. In the context of the Dymnikova black hole, quasinormal modes were investigated in \cite{Macedo:2024dqb}. 

Another important feature intensively studied in the last years is the presence of a quintessence field around a black hole, motivated by the accelerated expansion of the universe which is explained as being caused by the presence of a repulsive form of energy known as dark energy \cite{SupernovaCosmologyProject:1998vns,SupernovaSearchTeam:1998fmf,SDSS:2003eyi}. There is a wide range of models that attempt to explain this phenomenon; one of them is based on the introduction of a scalar field with a decreasing potential, called quintessence, which is responsible for influencing cosmological dynamics \cite{Carroll:1998zi}. In this work, we study the Dymnikova black hole surrounded by quintessence matter, following the model proposed by Kiselev. The study of black hole quasinormal modes and thermodynamics in the context of General Relativity, particularly in the presence of a quintessence field, has received significant interest and has been widely explored in previous research \cite{Chen:2005qh,Zhang:2006hh,Al-Badawi:2019tom,Pedraza:2021hzw,Sekhmani:2025zwc}.

In this work, we investigate the properties of a Dymnikova BH immersed in a quintessential field, characterized by the state parameter $\omega$ and a normalization constant $c$. We explore the thermodynamic behavior, null geodesics, scalar quasinormal modes and shadow profiles for this model. The paper is organized as follows: In Sec. II we present the solution for Dymnikova black hole immersed by quintessence field. In Sec. III we discuss the thermodynamics associated with the black hole.  In Sec.IV provides a study about the null geodesics and black hole shadows. In Sec V we present the numerical results obtained for quasinormal frequencies. Finally, we summarize the paper in Sec VI.

\section{Dymnikova Black Hole}

The metric corresponding to the Dymnikova black hole is given by  \cite{Dymnikova:1992ux}
\begin{equation}\label{2.1}
    ds^2 = -f(r)dt^2 + \dfrac{dr^2}{f(r)} + r^2(d\theta^2 + \sin^2{\theta}\hspace{0.1cm}d\varphi^2),
\end{equation}
where
\begin{equation}\label{2.2}
    f(r) = 1 - \dfrac{r_g}{r}(1 - e^{-r^{3}/r_{*}^3})
\end{equation}
and $r_{*}^3 = r_g r_0^2$, with $r_g$ denoting the Schwarzschild radius and $r_0$ the de Sitter radius.  Conversely, Kiselev proposed that a quintessence field can be integrated into the framework of a black hole \cite{Kiselev:2002dx}. The stress-energy tensor of this field, $(T_\mu^{\nu})^{(Q)}$ where $Q$ stands for quintessence, has the following components:
\begin{equation}
    (T_{\varphi}^{\varphi})^{(Q)} = (T_{\theta}^{\theta})^{(Q)} = -\dfrac{1}{2}(3w + 1)(T_{r}^{r})^{(Q)} = -\dfrac{1}{2}(3w + 1)(T_{t}^{t})^{(Q)}
\end{equation}
and $w$ is the state parameter which satisfies the dominant energy condition $(T_{tt})^{(Q)} \geq 0$ and $|3w + 1|\leq 2$ \cite{Malakolkalami:2015cza}. Based on Kiselev's approach, we analyze the Dymnikova black hole immersed in a quintessence field. Adding the term $-cr_g/r^{3w + 1}$ into the black hole metric \cite{Pedraza:2020uuy,Saleh:2018hba}, we get
\begin{equation}\label{2.4}
    f(r) = 1 - \dfrac{r_g}{r}(1 - e^{-r^{3}/r_{*}^3}) - \dfrac{r_gc}{r^{3w + 1}}.
\end{equation}
 where $c$ denotes a normalization factor associated with the quintessence matter field and $\omega$ has the range $-1<\omega<-1/3$ so as to stay aligned with astronomical evidence of an expanding universe in acceleration \cite{SupernovaCosmologyProject:1998vns,SupernovaSearchTeam:1998cav}. For $c \rightarrow 0$, we recover the Dymnikova black hole, for $r \gg r_{*}$ we find the  Schwarzschild black hole solutions  surrounded by quintessential matter obtained by Kiselev \cite{Kiselev:2002dx} and for $r\ll r_{*}$  we get the de Sitter black hole surrounded by quintessence  \cite{Liu:2018ypo}.
\begin{figure*}[ht!]
    \centering
    \begin{subfigure}{0.5\textwidth}
    \includegraphics[scale=0.31]{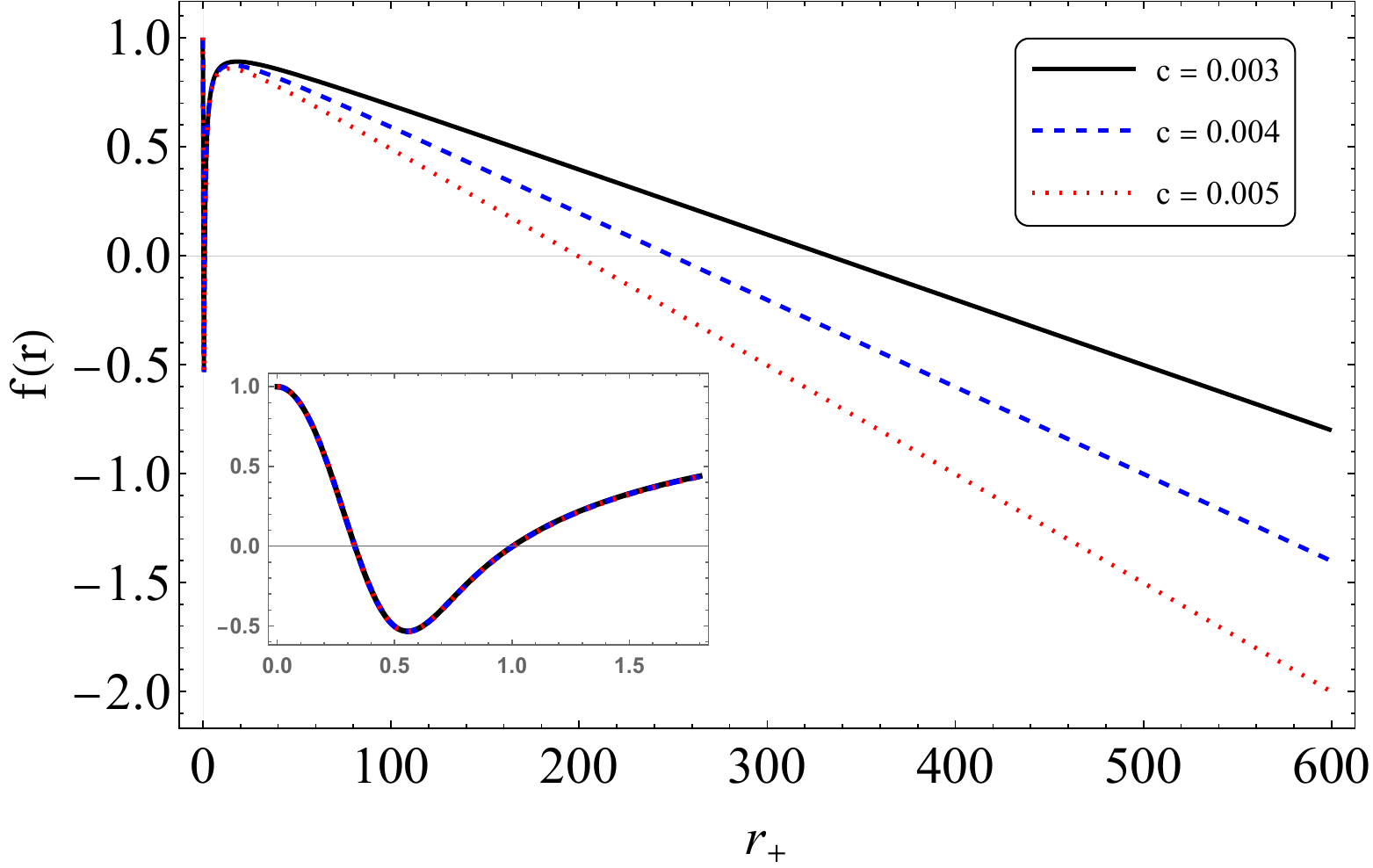}
    \subcaption{$(r_g, r_0, w) = (1, 0.3, -2/3)$  }
  \end{subfigure}%
  \begin{subfigure}{0.5\textwidth}
    \includegraphics[scale=0.31]{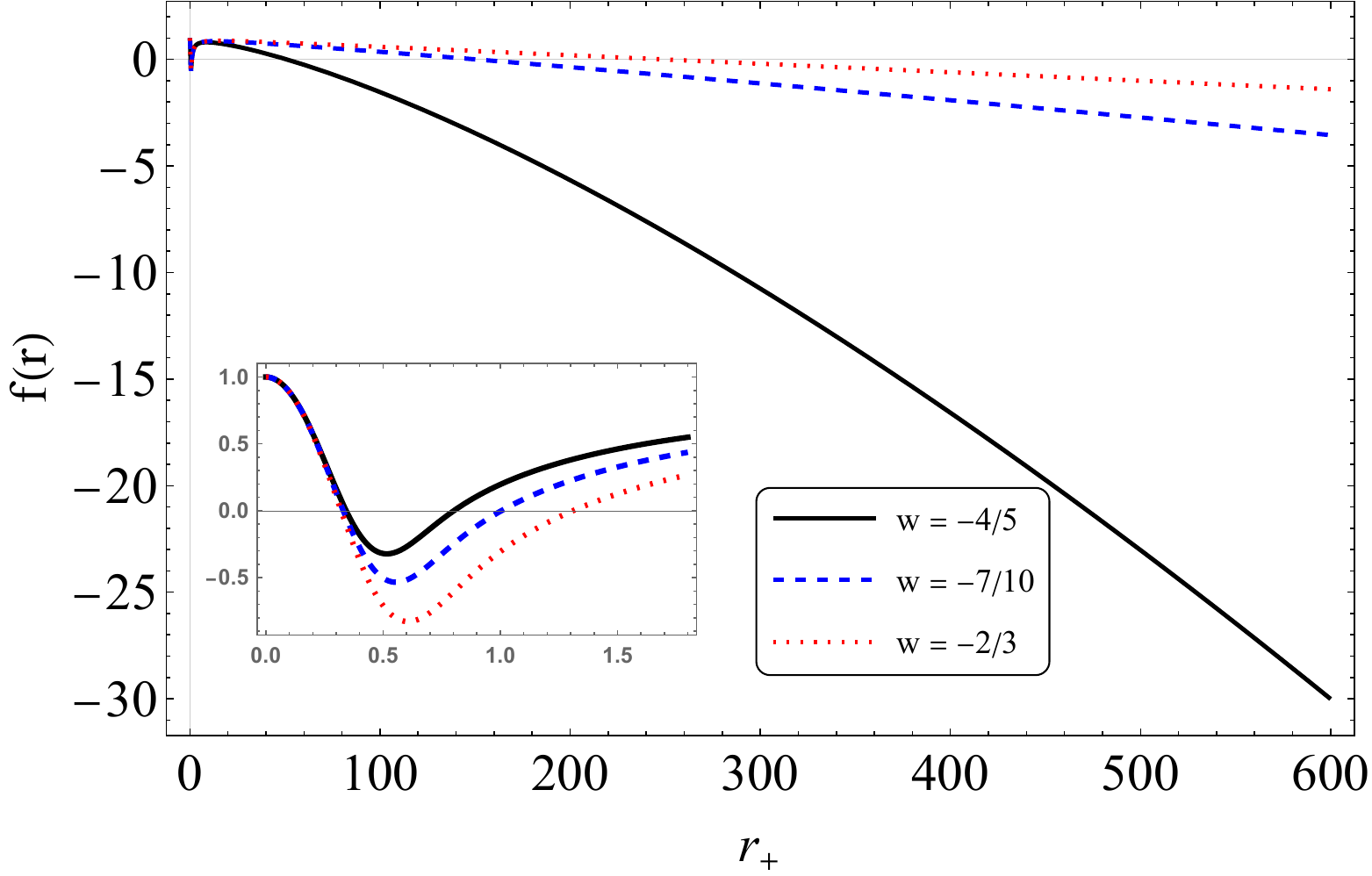}
     \subcaption{$(r_g, r_0, c) = (1, 0.3, 0.004)$  }
  \end{subfigure}
  \begin{subfigure}{0.5\textwidth}
    \includegraphics[scale=0.31]{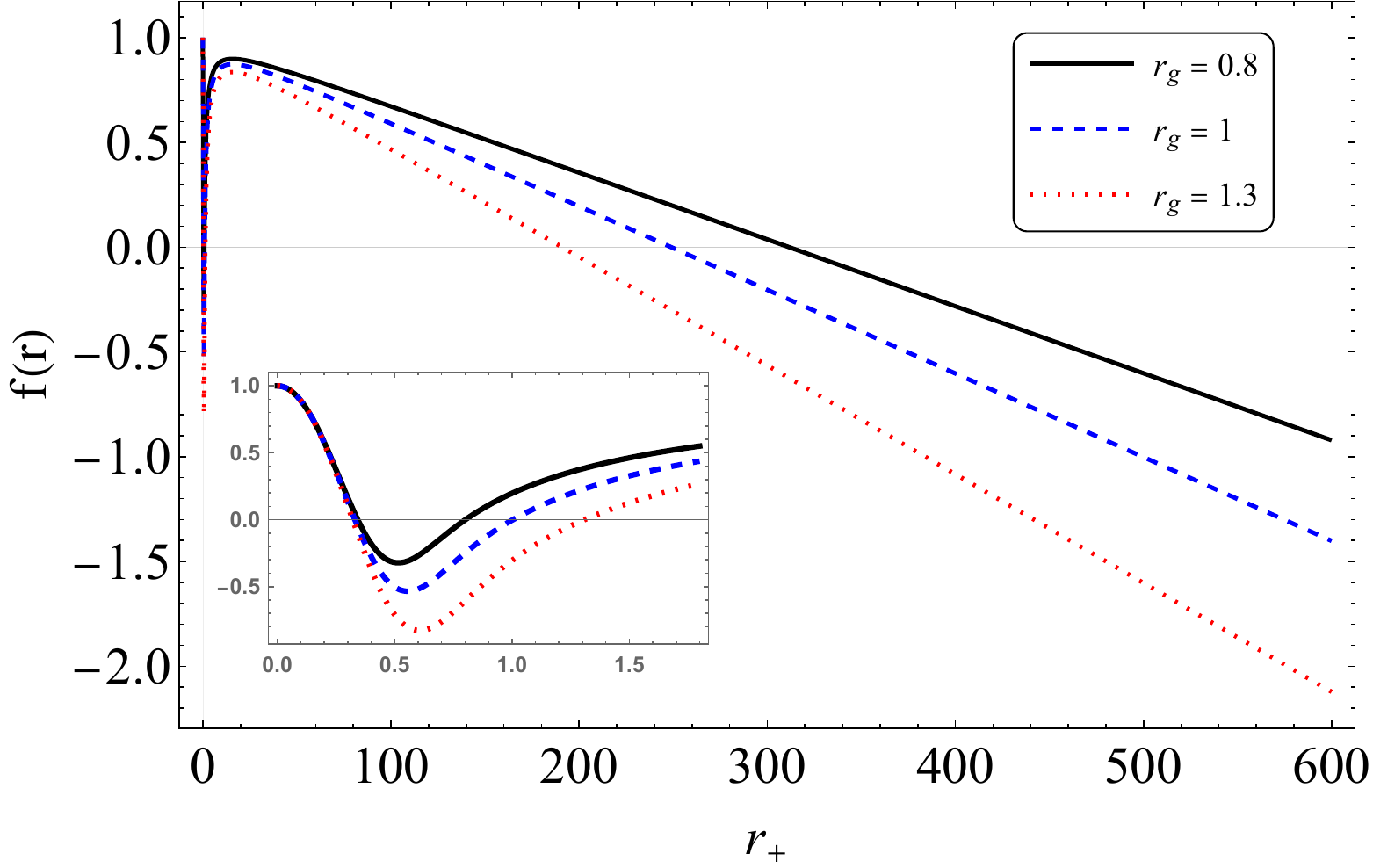}
     \subcaption{$(r_0, c, w) = (1, 0.004, -2/3)$  }
  \end{subfigure}%
  \begin{subfigure}{0.5\textwidth}
    \includegraphics[scale=0.31]{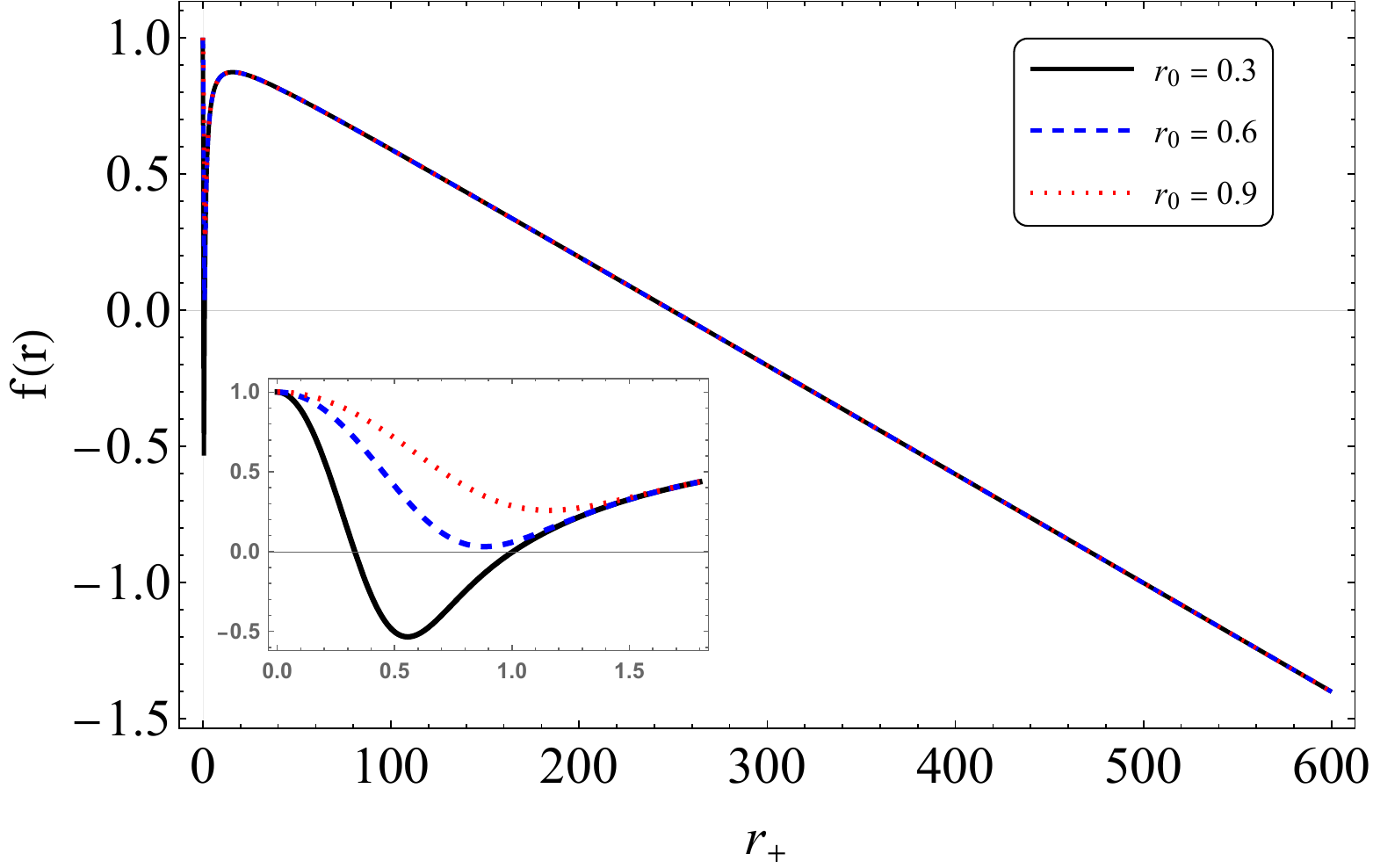}
     \subcaption{$(r_g, b, w) = (1, 0.004, -2/3)$  }
  \end{subfigure}
  \caption{The function $f(r)$ is plotted for various sets of the parameters $c, r_0, r_g$ and $w$.}
  \label{Figure 1}
\end{figure*}

When quintessence is added to the Dymnikova black hole, the energy–momentum tensor still describes an anisotropic, spherically symmetric vacuum; that is, it satisfies the relation:
 \begin{equation}
     (T\indices{_0^0})^{(DQ)} = (T\indices{_1^1})^{(DQ)}, \qquad \qquad (T\indices{_2^2})^{(DQ)} = (T\indices{_3^3})^{(DQ)}
 \end{equation}
where $DQ$ stands for Dymnikova with quintessence. The strong energy condition (SEC) remains violated for sufficiently large $r$ although one can obtain a region  at moderately small $r$ where the SEC is satisfied, depending on the quintessence parameters, $r_g$ and $r_0$.

Some recent works have employed this approach \cite{Chen:2022dap,Toshmatov:2015npp,Belhaj:2020rdb}. To determine the horizons, we must solve $f(r) = 0$. However, analytical expressions for the horizons cannot be obtained. The plot of equation (\ref{2.4}) is depicted in Fig.\ref{Figure 1} for various values of $c, r_0, r_g$ and $w$. As we can see, the number of horizons is strongly dependent on the parameter values and that the first two horizons, the inner and event, are quite close to each other, while the third one,  the cosmological, is significantly farther away from them. We observe that the parameters $c$ and $w$ do not influence the occurrence of the horizon for small values of $r$, but they do affect its location at large values of $r$. In contrast, the variation of $r_g$ and $r_0$ significantly alters the location of the horizon at small $r$ even to the extent that the horizon may cease to exist. However, as $r$ increases, the horizon remains unchanged. 

To determine the existence of curvature singularities, it is sufficient to examine the Kretschmann scalar, $K = R^{\mu\nu\sigma\rho}R_{\mu\nu\sigma\rho}$ due to the spherical symmetry of spacetime, which is equal to
\begin{equation}
    \begin{aligned}
    K  
    & = 4\frac{R_g^2(r)}{r^6} + 4\left(\frac{3}{r_0^2}e^{-\frac{r^3}{r_{*}^3}} - \frac{R_g(r)}{r^3}\right)^2 + \left(\frac{2R_g(r)}{r^3} - \frac{9r^3}{r_0^4r_g}e^{-\frac{r^3}{r_{*}^3}}\right)^2 \\& + \frac{3c^2r_g^2\{4 + w\{20 + 3w[17 + 9w(2+w)\} \}}{r^{6(1+w)}}  \\&
     + \frac{2cr_g}{r^{3(2 + w)}}\left\{6(1 + w)(2 + 3w)R_g(r) + \frac{3r^3(1 + 3w)}{r_0^2}e^{-\frac{r^3}{r_{*}^3}}\left[-3(4 + 3w)  + (2 + 3w) \left(2 - \frac{9r^3}{r_0^4r_g} \right) \right] \right\}
\end{aligned}
\end{equation}
This allows us to conclude that Dymnikova's spacetime is not regular anymore when quintessence is included.

\section{Thermodynamics}

We begin this section by exploring the thermodynamics of the Dymnikova black hole surrounded by quintessence matter. To achieve this, we first analyze the Hawking temperature
\begin{equation}
    T = \dfrac{\kappa}{2\pi} = \dfrac{1}{4\pi}\left.\dfrac{df(r)}{dr}\right|_{r = r_h}
\end{equation}
with $f(r)$ given by Eq.(\ref{2.4}) we derive the modified Hawking temperature as  
\begin{equation}
    T = \dfrac{1}{4\pi r_0}\left[\dfrac{r_0}{r_h} - \dfrac{3r_h}{r_0}\left(1 - \dfrac{r_h}{r_g}\right) - \dfrac{3cr_g r_0}{ r_h^{3w + 2}}\left( \dfrac{r_h^3}{r_g r_0^2} - w \right) \right]. 
\end{equation}
\begin{figure*}[ht!]
    \centering
    \begin{subfigure}{0.5\textwidth}
        \includegraphics[scale=0.41]{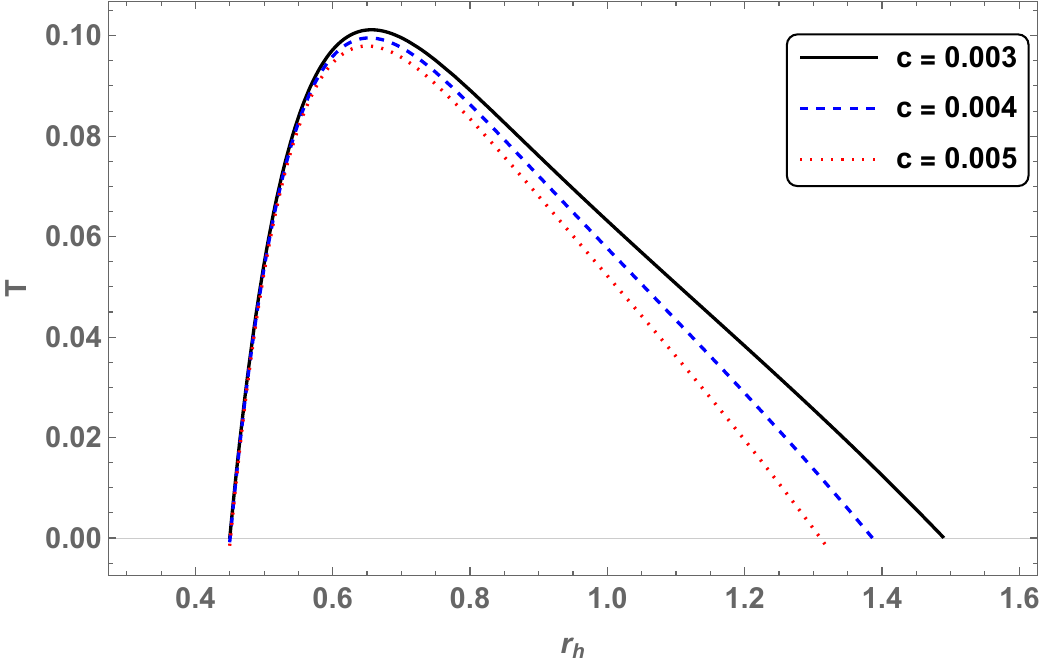}
        \subcaption{$(r_0, w) = (0.3, -2/3)$  }
    \end{subfigure}%
  \begin{subfigure}{0.5\textwidth}
    \includegraphics[scale=0.41]{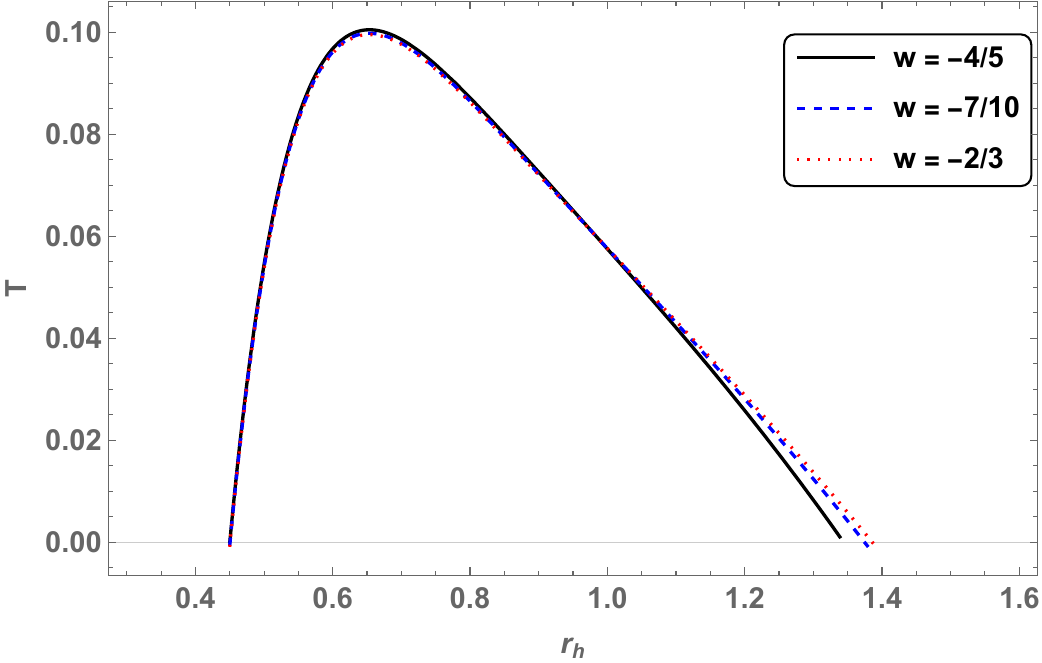}
     \subcaption{$(r_0, c) = ( 0.3, 0.004)$  }
  \end{subfigure} 
  \begin{subfigure}{0.5\textwidth}
    \includegraphics[scale=0.41]{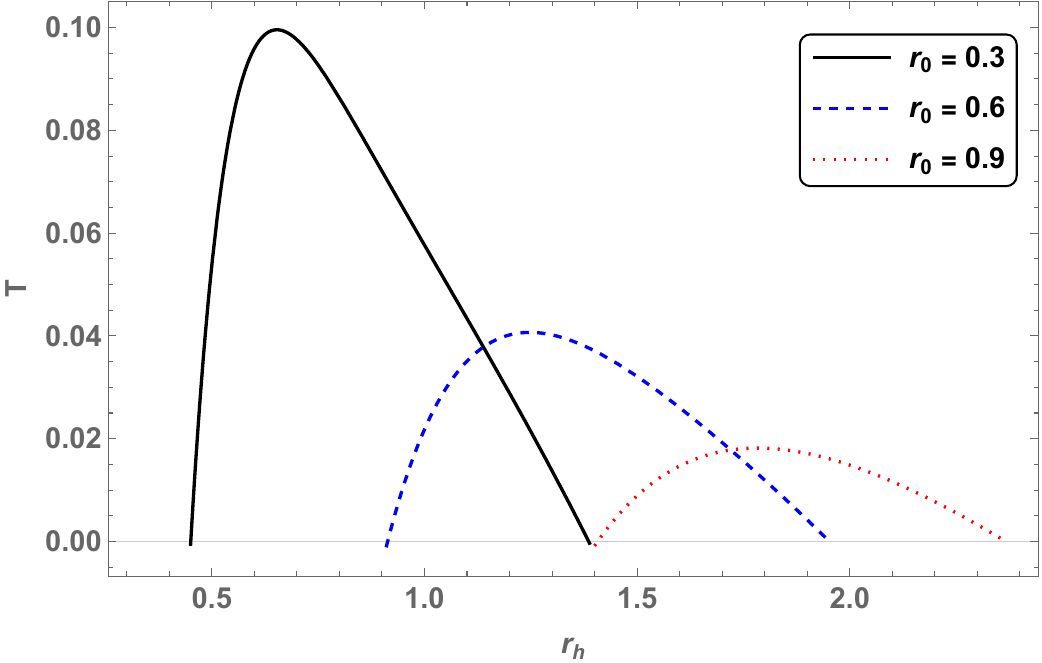}
     \subcaption{$(c, w) = (0.004, -2/3)$  }
  \end{subfigure}
  \caption{The Hawking temperature $T$ is plotted for various sets of the parameters $c, r_0, r_g$ and $w$ as function of $r$.}
  \label{Figure 2}
\end{figure*}
Here $r_h$ is the horizon radius. In the case $c = 0 $, this reduces to the Hawking temperature of Dymnikova black hole \cite{Dymnikova:1996}.  The maxima of the temperature cannot be obtained analytically, however to visualize these results, we present in Fig.\ref{Figure 2} a plot of the Hawking temperature as a function of the event horizon. The graph shows that the temperature first rises as the horizon radius grows, peaks at a certain point, and then gradually declines as the horizon radius continues to expand. The peak temperature value decreases as $c$ increases, moreover the point where the temperature vanishes shifts to progressively smaller values of $r_{+}$. Also it is observed that the peak Hawking temperature decreases as the state parameter $c$ increases, although now the point where the temperature vanishes occurs at increasingly larger values of $r_{+}$ and the same behavior occurs for an increase in $r_0$, while the other parameters  remain  constant.

Setting $f(r_h) = 0$  leads us to obtain the mass of the black hole  given by
\begin{equation}
    M = \dfrac{r_h}{2}\left[1 - \dfrac{c}{r_h^{3w}} + \frac{r_0^2}{r_h^2}W\left( -\dfrac{r_h^2}{r_0^2}e^{-\frac{r_h^2}{r_0^2}(1 - c/r_h^{3w})}\right)  \right]^{-1}
\end{equation} 
where $W$ is the Lambert Function \cite{Lehtonen:2016}. The expression above reduces to the Dymnikova black hole mass \cite{Macedo:2024dqb} when $c = 0$.

The determination of potential phase transitions in the black hole hinges on the criterion for a change in specific heat capacity sign. A positive specific heat capacity ($C > 0$) is a signature of local stability against thermal fluctuations, whereas a negative specific heat capacity ($C < 0$) indicates local instability. The expression for specific heat capacity $C = \frac{dM}{dT}$ is as follows: 
\begin{equation}
    C = \dfrac{\frac{4\pi M r_0}{r_h}\left\{1 - \frac{ \left[\frac{3wc}{r_h^{3w}} - \dfrac{2r_0^2}{r_h^2}W\left(-\dfrac{r_h^2}{r_0^2}e^{-\frac{r_h^2}{r_0^2}\left(1 - \frac{c}{r_h^{3w}}\right) }  \right) + \dfrac{r_0^2}{r_h}\dfrac{\partial W}{\partial r}  \right] }{\left[1 - \dfrac{c}{r^{3w}} + \dfrac{r_0^2}{r_h^2}W\left(-\dfrac{r_h^2}{r_0^2}e^{-\frac{r_h^2}{r_0^2}\left(1 - \frac{c}{r_h^{3w}}\right) } \right) \right] } \right\}}{ \left\{-\dfrac{r_0}{r_h^2} - \dfrac{3}{r_0}\left(1 - \dfrac{2r_h}{r_g}\right)  + \dfrac{cr_g}{r_h^{3w+1}}\left[\dfrac{(3w + 2)}{r_h^2} + \dfrac{3(3w - 1) r_h }{r_g r_0^2} - \dfrac{(3w - 2)(3w + 1)}{r_h^2}     \right]  \right\} }
\end{equation}
\begin{figure*}[ht!]
    \centering
    \begin{subfigure}{0.5\textwidth}
    \includegraphics[scale=0.41]{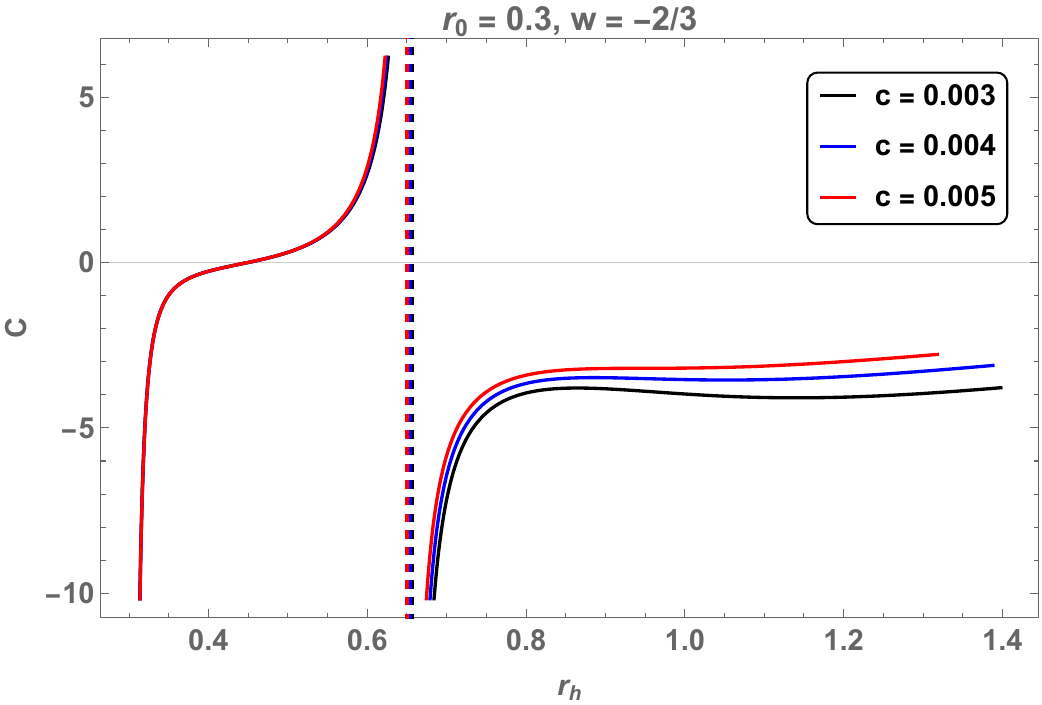}
    \subcaption{$r_+^{C} = 0.65$}
  \end{subfigure}%
  \begin{subfigure}{0.5\textwidth}
    \includegraphics[scale=0.41]{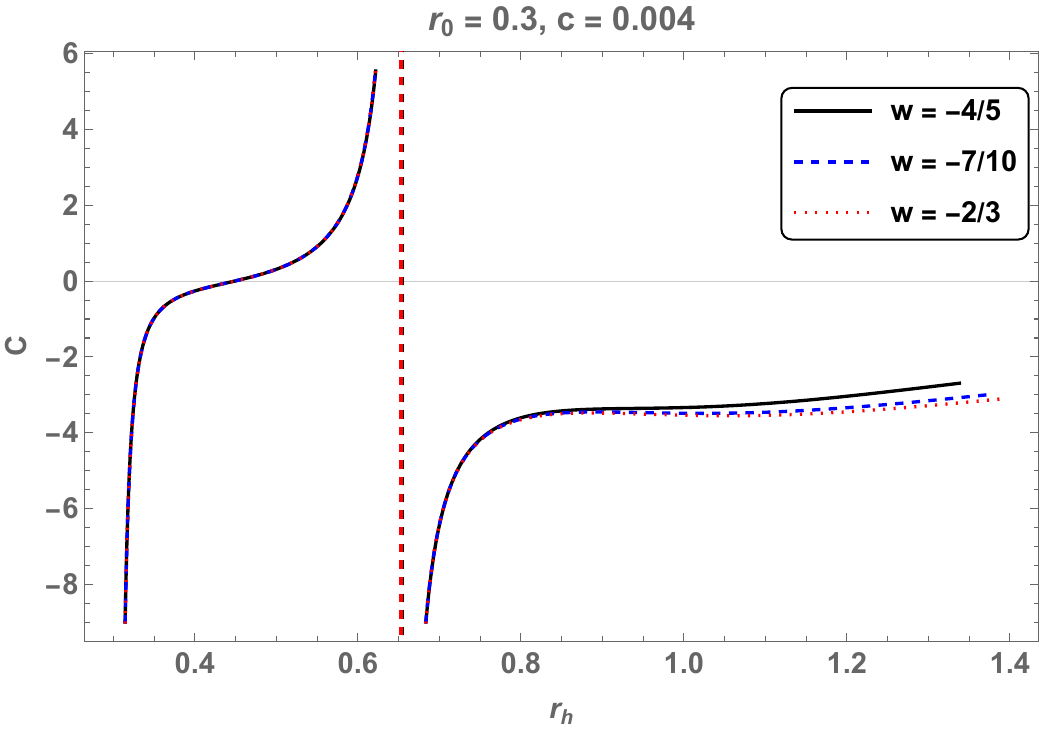}
     \subcaption{$r_{+}^{C} =  0.65$  }
  \end{subfigure}
  \begin{subfigure}{0.5\textwidth}
    \includegraphics[scale=0.41]{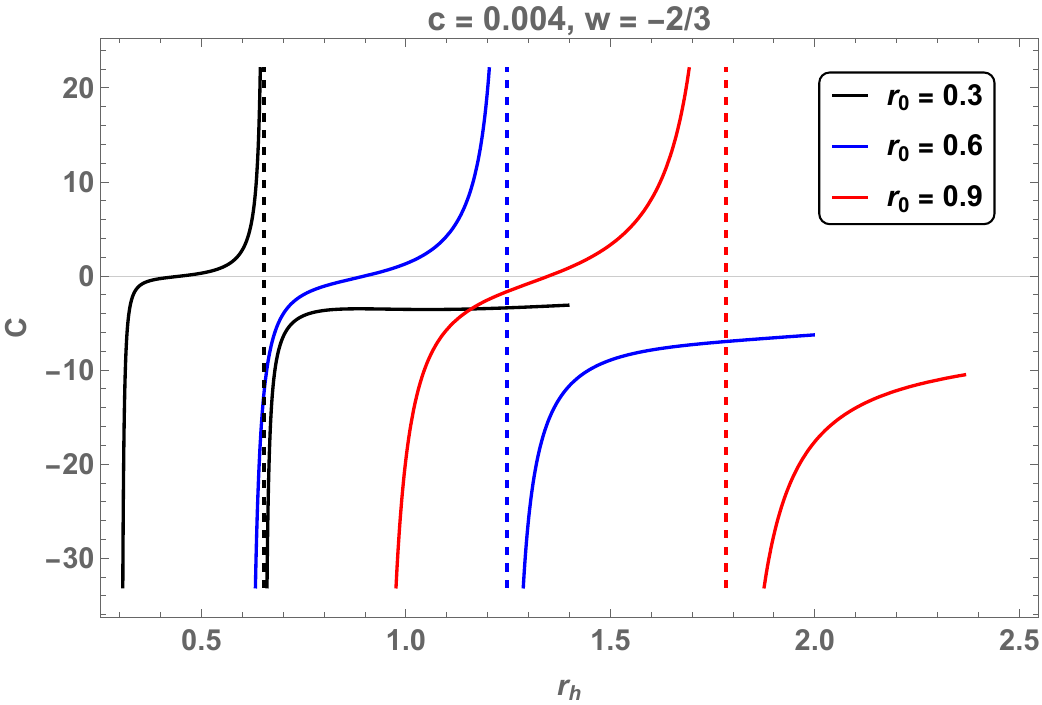}
     \subcaption{$(r_{+}^{C}$, \textcolor{blue}{$r_{+}^{C}$}, \textcolor{red}{$r_{+}^{C}$}) = (0.65, 1.24, 1.78)  }
  \end{subfigure}%
  \caption{The thermal capacity $C$ is plotted for various sets of the parameters $c, r_0, r_g$ and $w$ as function of $r$.}
  \label{Figure 3}
\end{figure*}

A numerical analysis shows that the specific heat capacity diverges at the point where the Hawking temperature reaches its maximum, changing from negative infinity to positive infinity. This point, namely $r_{+}^{C}$, represented by a vertical line in the graph, is known as the Davies point and marks the occurrence of the phase transition of the black hole. Let $r_{+}^{(0)}$ be the point where $T = C = 0$, then the heat capacity becomes positive in the region $r_{+}^{(0)} < r_{+} < r_{+}^{C}$. It is negative within the interval $0 < r_{+} < r_{+}^{(0)}$ and also for $r_{+} > r_{+}^{C}$. By plotting  the specific heat capacity $C$ with the horizon radius as shown in Fig \ref{Figure 3}, we see the dependence of the Davies point on the parameters $c, w$ and $r_0$. For quintessence parameter $c$ and for state parameter $w$, we observe that  the Davies point practically does not change. In contrast, from the context of dependence on $r_0$, the Davies points shifts toward larger horizon radius within increasing values of $r_0$.

\section{Geodesics}

\begin{figure*}[ht!]
    \centering
    \begin{subfigure}{0.5\textwidth}
    \includegraphics[scale=0.45]{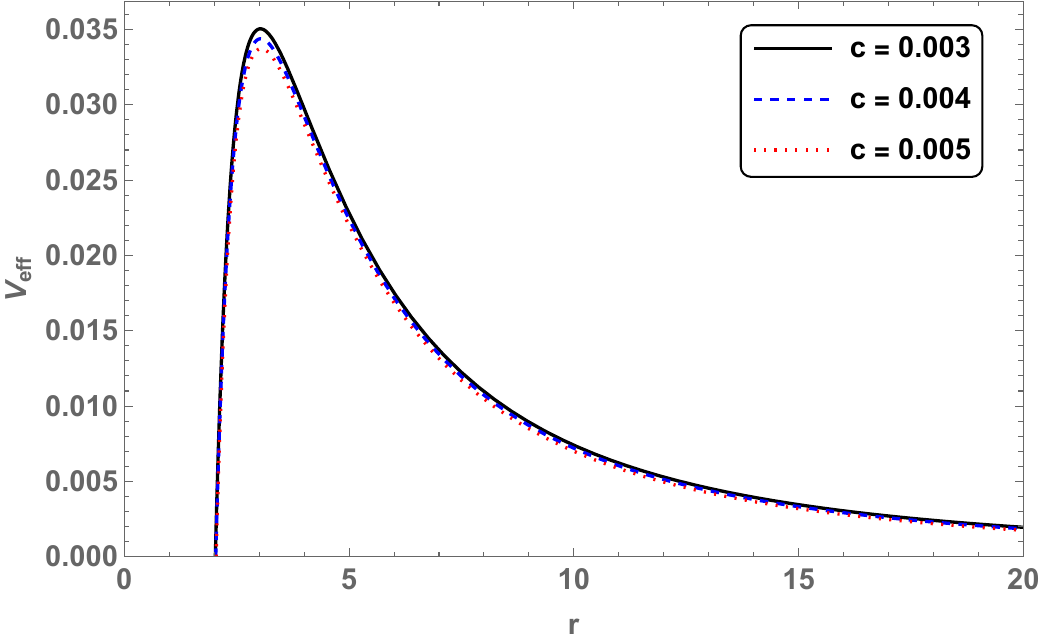}
    \subcaption{$(r_0, w) = (0.3, -2/3)$  }
  \end{subfigure}%
  \begin{subfigure}{0.5\textwidth}
    \includegraphics[scale=0.45]{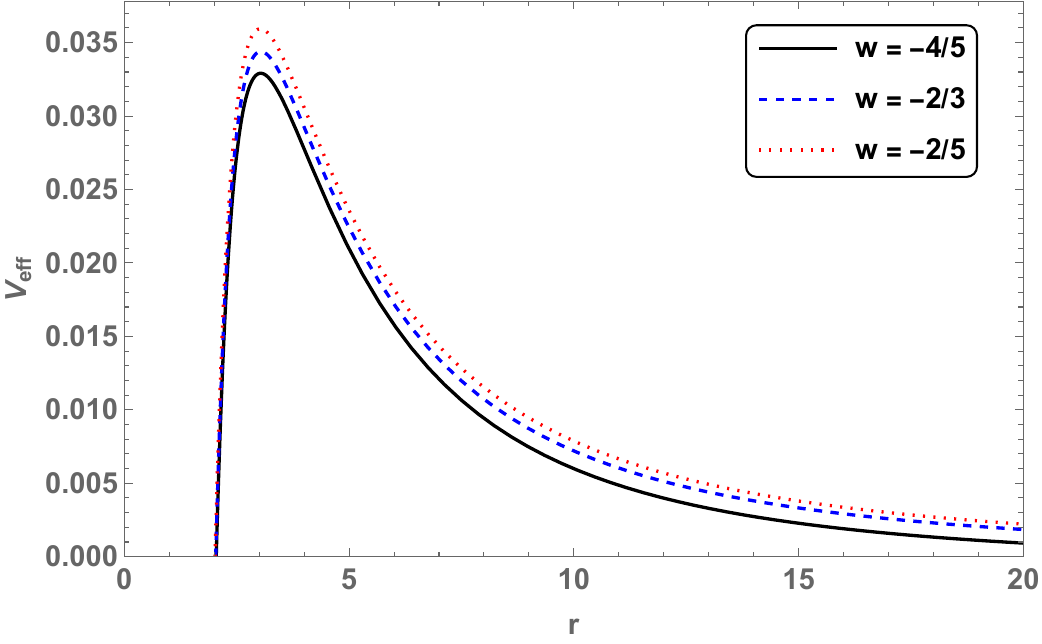}
     \subcaption{$(r_0, c) = (0.3, 0.004)$  }
  \end{subfigure}
  \caption{The effective Potential $V_{eff}$ of the black hole is plotted for various sets of the parameters $c$ and $w$ as function of $r$.}
  \label{Figure 4}
\end{figure*}

In this section, we will analyze the geodesic motion of test particles around Dymnikova black hole surrounded by quintessence. The metric (\ref{2.1}) is independent of the coordinates $t$ and $\varphi$, which in turn implies the existence of a Killing vector associated with translational symmetry and another one associated with rotational symmetry. Moreover, due to the spherical symmetry, it is possible to restrict the metric to the equatorial plane by fixing $\theta = \pi/2$. Therefore, the metric can now be written as:
\begin{equation}\label{4.10}
    ds^2 = -f(r)dt^2 + f(r)dr^2 + r^2d\varphi^2
\end{equation}
From the Killing vector $K_\mu = (-f(r), 0, 0, 0)$, we obtain the following conserved quantity:
\begin{equation}
    E = f(r)\dot{t}
\end{equation}
while for $K_\mu = (0, 0, 0, r^2)$, we find:
\begin{equation}\label{4.12}
    L = r^2\dot{\varphi}
\end{equation}
where `$.$' denotes the differentiation with respect to the affine parameter $\lambda$. The Lagragian function, $\mathcal{L} =\dfrac{1}{2}\dot{x}^\mu \dot{x}_\mu$, for metric (\ref{4.10})  is given by
\begin{equation}
    \mathcal{L} = \dfrac{1}{2}\left(-f(r)\dot{t}^2 + \dfrac{\dot{r}^2}{f(r)}  + r^2\dot{\varphi}^2 \right)
\end{equation}
The previous equation can be rewritten as:
\begin{equation}\label{4.14}
    \dot{r}^2 = E^2 - f(r)\left(\epsilon + \dfrac{L^2}{r^2} \right)
\end{equation}
where $\epsilon = -2\mathcal{L} $ represents the norm of the tangent vector along the geodesic path, and it is equal to $0$ and $1$ for null and time like geodesics, respectively. Therefore, 
\begin{equation}\label{4.15}
    \dot{r}^2 = E^2 - V_{eff}
\end{equation}
where 
\begin{equation}
    V_{eff} = \left[1 - \dfrac{r_g}{r}\left(1 - e^{-r^3/r_{*}^3}\right) - \dfrac{r_gc}{r^{3w + 1}} \right]\left(\epsilon + \dfrac{L^2}{r^2} \right)
\end{equation}
represents the effective potential for Dymnikova black hole surrounded by quintessence. The Fig.\ref{Figure 4} shows the behavior of $V_{eff}$ for null geodesics with different values of model parameters. In panel (a), the parameters $r_0$ and $w$ are fixed while $c$ is varied. An increase in $c$ leads to a slight decrease in the peak of the potential. In panel (b), $r_0$ and $c$ are fixed while $w$ varies. As $w$ increases, the peak of $V_{eff}$ also increases. From Eq.(\ref{4.12}) and (\ref{4.15}) we get the deflection of light rays near the black hole:
\begin{equation}
    \dfrac{d\varphi}{dr} = -\dfrac{b}{r^2}\left(1 - \dfrac{b^2}{r^2}f(r)\right)^{-1/2}
\end{equation}
where $b = L/E$ is the impact parameter. 

%Gráfico está no arquivo Cálculo Geométrico.nb%

\begin{figure}[h!]
    \centering
    \includegraphics[scale=0.53]{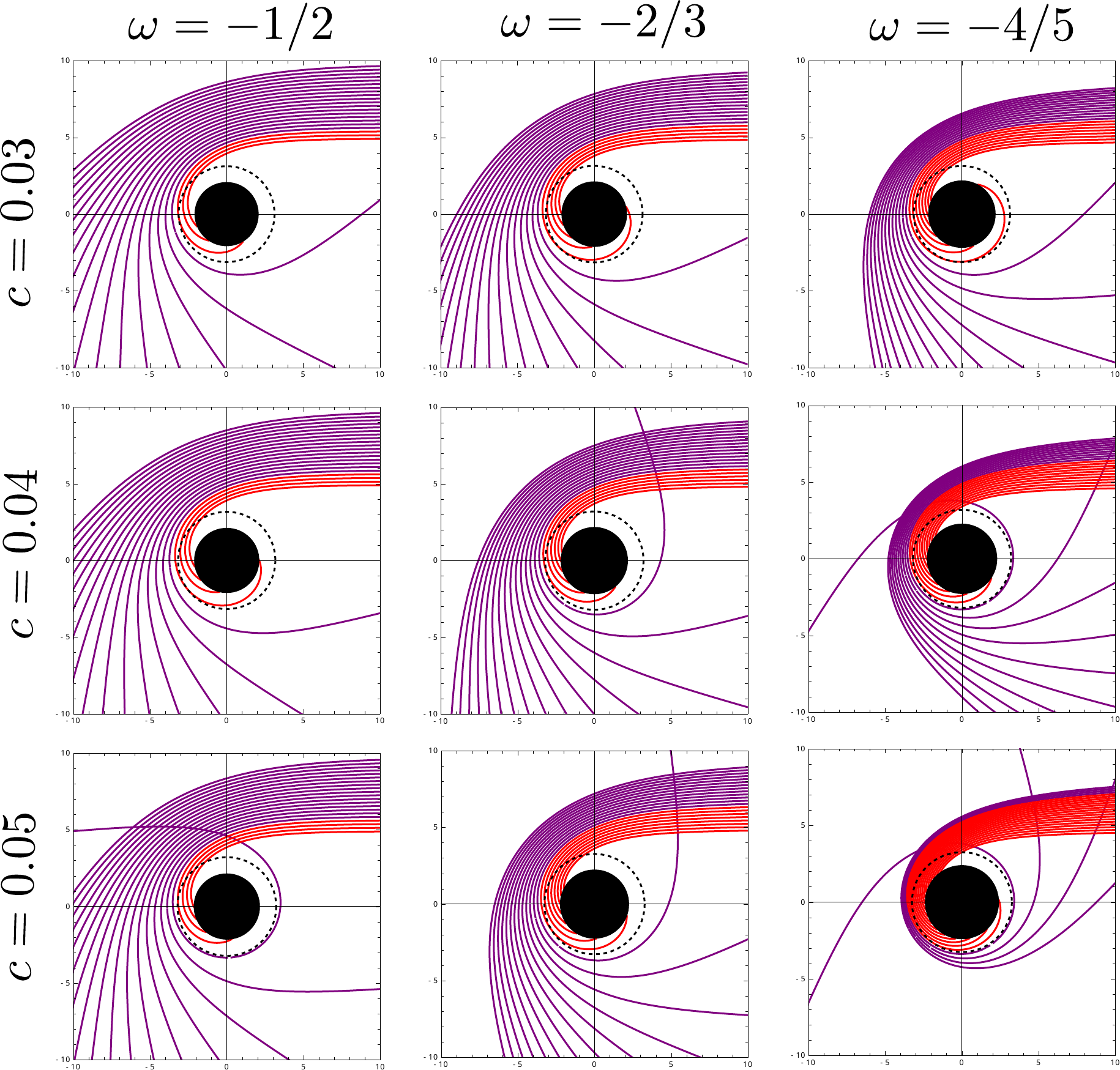}
    \caption{Geodesics of massless particles near the Dymnikova black hole surrounded by quintessence. The purple lines indicates the particles that do not fall in the black hole, going therefore to cosmological horizon, while the red lines are for those particles that end up falling in the black hole. For this plot we have considered $M=1$ and $r_0=0.3$.}
    \label{Figure geodesic}
\end{figure}

To better understand the physical behavior of light rays, Fig. \ref{Figure geodesic} illustrates how their paths change depending on the model parameters. It is evident that each parameter significantly affects the behavior of light rays, altering the degree of deflection along their trajectories.  Evidently, for values greater than the critical impact parameter, the light rays are deflected by the black hole but eventually escape to cosmological horizon. On the other hand, for values less than the critical impact parameter, the rays spiral into the black hole, crossing the event horizon.

In summary, Fig. \ref{Figure geodesic} presents the geodesics of massless particles near the Dymnikova black hole surrounded by quintessence. The purple lines indicate the particles that do not fall in the black hole, going therefore to cosmological horizon, while the red lines indicate the particles that end up falling in the black hole. For this plot, we have considered $M=1$ and $r_0=0.3$. The black disk represents the black hole itself, limited by the event-horizon radius, and the dashed circle stands for the photon ring. Notice that all the particles that cross the dashed circle end up falling in the black hole. Crossing the limit of the photon ring means crossing the peak of the effective potential related to an unstable equilibrium of light-like particles. Also notice that, for any given value of $c$, as the $\omega$ parameter gets smaller the black hole tends to enhance the confinement of the particles. The opposite behavior occurs if we consider a fixed $\omega$. In this case, the increasing values of the parameter $c$ intensify the confinement of the particles. Such geodesics behaviors are directly related to the increasing/decreasing of the horizon radius for the considered configurations, as we can see from table \ref{table_radius}. It is important to highlight here that, for all the plots, we have considered the same range for the impact parameter of the ingoing particles.    

\subsection{Shadows}

\begin{figure*}[ht!]
    \centering
\begin{subfigure}{\textwidth}
    \centering
    \begin{subfigure}{0.48\textwidth}
        \includegraphics[width=\textwidth]{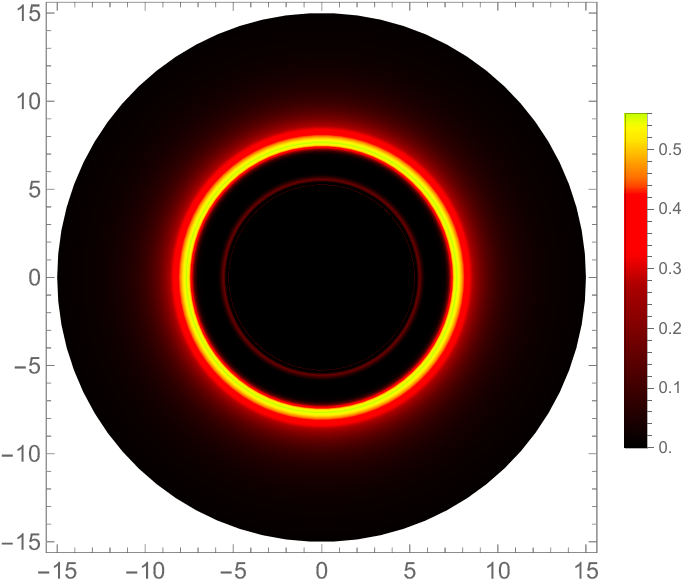}
    \end{subfigure}
    \hfill
    \begin{subfigure}{0.48\textwidth}
        \includegraphics[width=\textwidth]{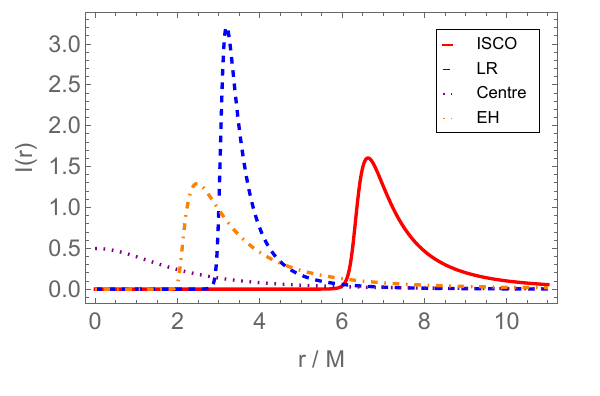}
    \end{subfigure}
    \subcaption{$c = 0.002$}
\end{subfigure}
\vspace{0.5em}
\begin{subfigure}{\textwidth}
    \centering
    \begin{subfigure}{0.48\textwidth}
        \includegraphics[width=\textwidth]{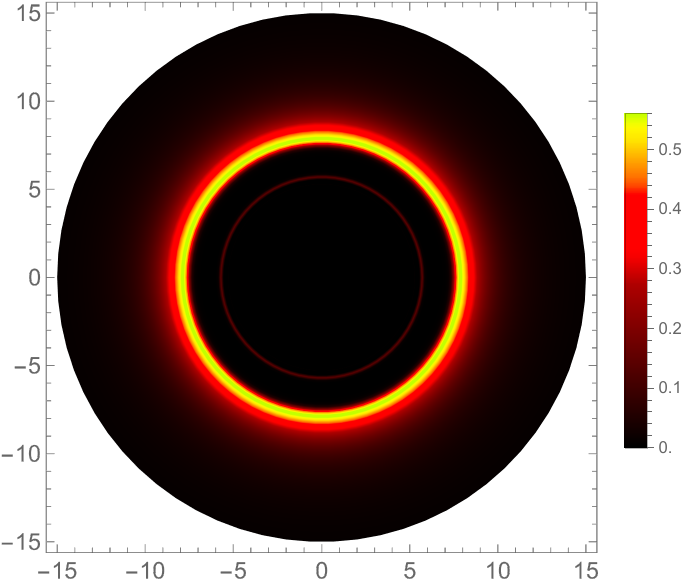}
    \end{subfigure}
    \hfill
    \begin{subfigure}{0.48\textwidth}
        \includegraphics[width=\textwidth]{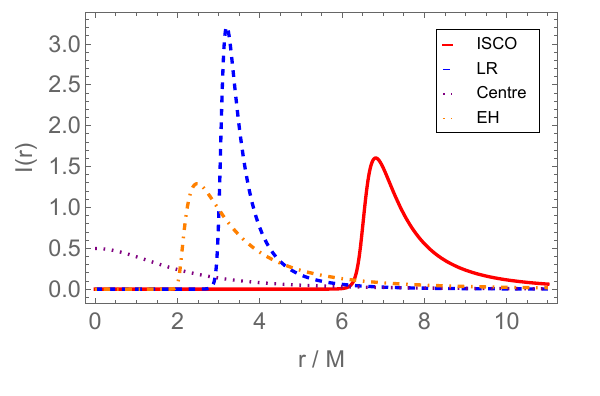}
    \end{subfigure}
    \subcaption{$c = 0.003$}
\end{subfigure}
\caption{Shadows and intensity profiles of the black hole for different values of the parameter $c$, with fixed $(r_0, w, M, L) = (0.3, -2/3, 1, 1)$.}
\label{Figure shadow}
\end{figure*}

 We begin by examining the influence of the accretion disk on the observed black hole shadow and the associated photon rings. The ray-tracing simulations are performed using a Mathematica-based code that has been utilized in previous studies \cite{Guerrero:2021ues, Guerrero:2022qkh, Rosa:2023qcv}. In this problem, the accretion disk is modeled as an infinitesimally thin structure confined to the equatorial plane $\theta = \pi/2$. The emission profile of the accretion disk, as a function of the radial coordinate, is described using the GLM model, which employs a Johnson SU distribution and  reproduce intensity profiles obtained from simulations of accretion disks in realistic astrophysical scenarios \cite{Gralla:2020srx, Vincent:2022fwj}. In this model, the intensity profile in the emitter's frame, $I$, is equal to
 \begin{equation}
     I(r, \mu, \sigma,\gamma) = \dfrac{\operatorname{exp}\left\{-\frac{1}{2}\left[\gamma + \operatorname{arcsinh}\left(\frac{r - \mu}{\sigma} \right) \right]^2 \right\} }{\sqrt{(r - \mu)^2 + \sigma^2}}
 \end{equation}
 The parameters $\mu, \sigma,\gamma$ are responsible for the location of the central peak, the overall width of distribution and the rate of increase, respectively. By adjusting these parameters, we can shape the insensity profile to reflect different astrophysical scenarios and explore how they affect the observed emission. In this work, we use four different intensity profiles: the ISCO model is defined by the parameter values $\gamma = -2, \mu = r_{ISCO}$, and $\sigma = M/4$, the light ring model characterized by the parameters $\gamma = -2, \mu = r_{LR}$ and $\sigma = M/8$, the center model adopts  $\gamma  =\mu = 0$ with $\sigma = r_{EH}$. Lastly, the event horizon model is described by the parameters $\gamma = -3$, $\mu  = r_{EH}$ and $\sigma = M/8$. The observed intensity of photons at a single frequency is defined as follows:
 \begin{equation}
     I_{obs} = f(r)^{3/2}I = \left[1 - \dfrac{r_g}{r}(1 - e^{-r^{3}/r_{*}^3}) - \dfrac{r_gc}{r^{3w + 1}}\right]^{3/2}I
 \end{equation}
 Figure \ref{Figure shadow} compares the shadows of the black hole for different values of the parameter $c_0$, with fixed $r_0 = 0.3, w = -2/3$ and $M = L = 1$.

\begin{table}
    \centering
    \begin{tabular}{c||c|c|c}\hline\hline
         & $\omega=-1/2$ & $\omega=-2/3$ & $\omega=-4/5$\\\hline\hline
        $c=0.03$ & 
        \begin{tabular}{c}
       $r_h=2.0906$ \\
       $r_{pr}=3.1242$ \\
       $r_{sh}=5.6403$ \\
        \end{tabular} & 
        \begin{tabular}{c}
       $r_h=2.1370$ \\
       $r_{pr}=3.1487$ \\
       $r_{sh}=6.0556$ \\
        \end{tabular} & 
        \begin{tabular}{c}
        $r_h=2.1987$ \\
        $r_{pr}=3.1402$ \\
        $r_{sh}=6.7848$ \\
        \end{tabular}\\\hline
         $c=0.04$& 
         \begin{tabular}{c}
        $r_h=2.1238$ \\
        $r_{pr}=3.1692$ \\
        $r_{sh}=5.8082$ \\
        \end{tabular} & 
        \begin{tabular}{c}
        $r_h=2.1922$ \\
        $r_{pr}=3.2055$ \\
        $r_{sh}=6.4387$ \\
        \end{tabular} & 
        \begin{tabular}{c}
        $r_h=2.2931$ \\
        $r_{pr}=3.1949$ \\
        $r_{sh}=7.7345$ \\
        \end{tabular}\\\hline
        $c=0.05$ & 
        \begin{tabular}{c}
        $r_h=2.1585$ \\
        $r_{pr}=3.2163$ \\
        $r_{sh}=5.9880$ \\
        \end{tabular} & 
        \begin{tabular}{c}
        $r_h=2.2540$ \\
        $r_{pr}=3.2667$ \\
        $r_{sh}=6.8956$ \\
        \end{tabular} & 
        \begin{tabular}{c}
        $r_h=2.4148$ \\
        $r_{pr}=3.2547$ \\
        $r_{sh}=9.2205$ \\
        \end{tabular}\\\hline\hline
    \end{tabular}
    \caption{Table with the values of the horizon radius ($r_h$), the photon ring radius ($r_{pr}$) and the shadow radius ($r_{sh}$) for the configurations of $(\omega, c)$ according to the panel of \ref{Figure geodesic}.}
    \label{table_radius}
\end{table}

  The angular shadow size is given by
    \begin{equation*}
        \theta_{shadow} = \dfrac{b_c}{D}
    \end{equation*}
    where $D$ is the distance to the observer and $b_c$ is the critical impact parameter.
    For Sgr $A^{*}$, we adopt $M = 4.3\times10^6 M_{\odot}$ and $D = 8.3 \operatorname{kpc}$ \cite{Gillessen:2009}. For the Dymnikova model, we use \cite{Dymnikova:2019vuz, Cardenas:2021eri}:
    \begin{equation*}
        r_0  =  2.4\times 10^{-27}\operatorname{m}, \qquad \qquad c = 10^{-12}\operatorname{m}^{-2} \qquad \qquad w = -2/3.
    \end{equation*}
     This yields: 
    \begin{table}
        \centering
        \begin{tabular}{c||c|c}
            \hline \hline
             & $b_{c}(\operatorname{m}) $ & $\theta_{shadow} (\mu as) $ \\ \hline \hline
            Dymnikova & $3.310\times 10^{10}$ & $26.670$ \\ \hline
            Schwarzschild & $3.308\times10^{10}$ & $26.655 $ \\ \hline \hline
        \end{tabular}
        \caption{Critical impact parameter $b_c$ and the shadow angular size $\theta_{shadow}$  for Dymnikova and Schwarzschild black holes.}
        \label{table2} 
    \end{table}   
    The EHT measurement for Sgr $A^{*}$ is  $25.9\pm 1.15$ $\mu as$. Therefore, the Dymnikova model lies well within the reported experimental uncertainty, showing no statistically significant deviation from the observational value.

\section{Quasinormal modes}

The wave equation for a massless scalar field in a curved spacetime can be expressed as
 \begin{equation}\label{5.13}
     \dfrac{1}{\sqrt{-g}}\partial_\mu(\sqrt{-g}\hspace{0.1cm} g^{\mu\nu}\partial_\nu)\Psi = 0
 \end{equation}
where $\Psi$ denotes the wave function associated with the scalar field.

\begin{figure*}[ht!]
    \centering
    \begin{subfigure}{0.5\textwidth}
    \includegraphics[scale=0.45]{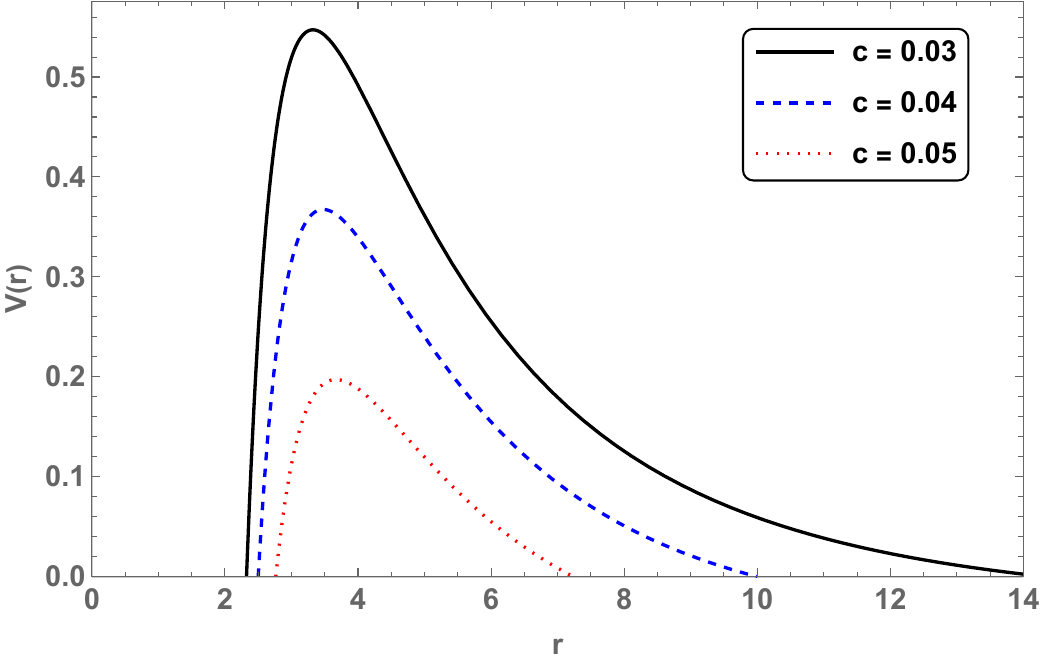}
    \subcaption{$(r_0, w) = (0.3, -2/3)$  }
  \end{subfigure}%
  \begin{subfigure}{0.5\textwidth}
    \includegraphics[scale=0.45]{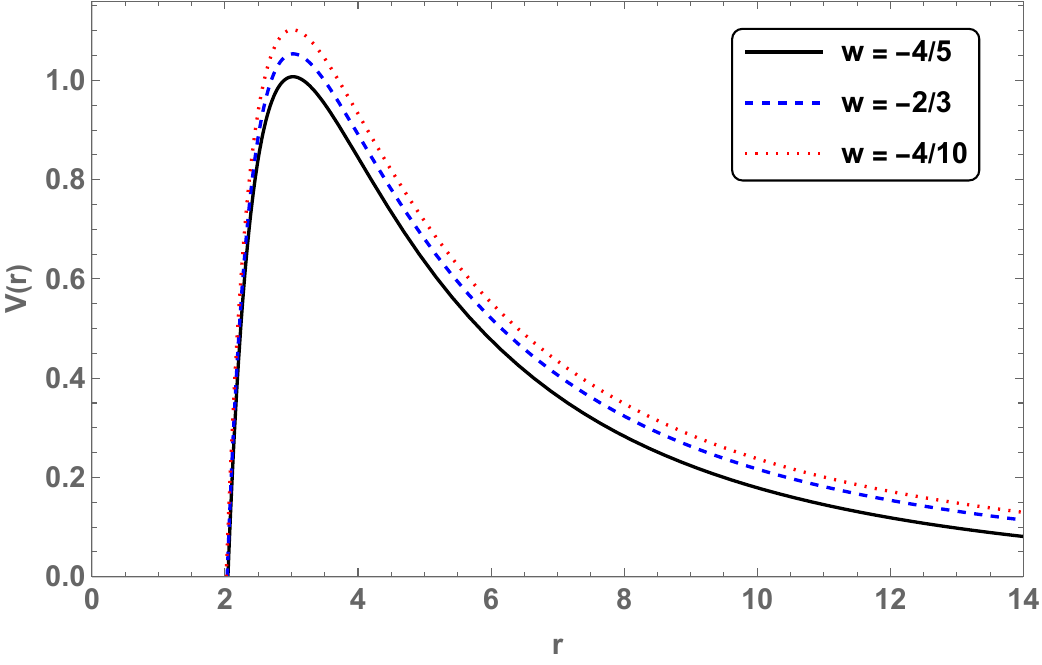}
     \subcaption{$(r_0, c) = (0.3, 0.004)$  }
  \end{subfigure}
  \caption{The Potential $V(r)$ of the black hole is plotted for various sets of the parameters $c$ and $w$ as function of $r$.}
  \label{Figure 7}
\end{figure*}

Using the separation of variables method, we assume the following ansatz for the scalar field \cite{Higuchi:1986}:
\begin{equation}\label{5.14}
    \Psi(t, r,\theta, \varphi) = e^{-i\omega t}Y(\theta,\varphi)\dfrac{\Phi(x)}{x}
\end{equation}
here $x$ is the tortoise coordinate transformation given by:
\begin{equation}
    x = \int \dfrac{dr}{1 - \dfrac{r_g}{r}(1 - e^{-r^3/r_{*}^3}) - \dfrac{r_gc}{ r^{3w + 1} } }
\end{equation}
and $Y(\theta,\varphi)$ denotes the spherical harmonics. As $x$ approaches negative infinity, it signifies the presence of the event horizon, whereas the limit as $x$ approaches positive infinity corresponds to the cosmological horizon. The QNM boundary conditions for the wave can be written in a compact form as follow \cite{Chrysostomou:2023jiv, Hod:2018fet, Dias:2020ncd}:
\begin{equation}
    \Psi \sim 
    \begin{cases}
        e^{-ik x}, & r \to r_+ \quad (x \to -\infty), \\[2pt] 
        e^{ikx}, & r\to r_c \quad (x\to +\infty).
    \end{cases}
\end{equation}
At the event horizon, it is reasonable to require that no waves are emitted, while at cosmological horizon it is assumed that no radiation is incoming from outside. The wave number $k$, which is greater than zero, satisfies the dispersion relations. Substituting the Eq.(\ref{5.14}) into Eq.(\ref{5.13}) we arrive at the radial wave equation in a Schrödinger-like form:
\begin{equation}
    \dfrac{d^2\Phi}{dx^2} + (\omega^2 - V)\Phi(r) = 0 
\end{equation}
with the effective perturbative potential, $V(r)$, given by
\begin{equation}
    V(r) = \frac{1}{r^2}\left[\ell(\ell + 1) + \frac{r_g}{r}(1 - e^{-\frac{r^3}{r_{*}^3}}) - \frac{3r^2}{r_0^2}e^{-\frac{r^3}{r_{*}^3}} + \frac{r_gc(3w+1)}{r^{3w + 1}} \right]\left[1 - \frac{r_g}{r}(1 - e^{-\frac{r^3}{r_{*}^3}}) - \frac{r_gc}{ r^{3w + 1}} \right]
\end{equation}
where $\ell$ is the orbital angular momentum. The potential of scalar field is plotted in Fig.\ref{Figure 7} for $r_g = 2, r_0 = 0.3, \ell = 5$ and different values of $c$ and $w$. The figure on the left shows that the height of the potential peak decreases as $c$ increases, the graph on the right display similar behavior, the peak decreasing as $w$ increases.
Using the WKB method, one obtains a closed-form expression for the quasinormal frequencies as \cite{Gogoi:2023lvw}:
\begin{equation}
    \begin{aligned}
    \omega^2 & = V_0 + A_2(\mathcal{K}^2) + A_4(\mathcal{K}^2) + A_6(\mathcal{K}^2) 
    - i\mathcal{K}\sqrt{-2V_0''}(1 + A_3(\mathcal{K}^2) + A_5(\mathcal{K}^2) + A_7(\mathcal{K}^2) + \ldots)
\end{aligned}
\end{equation}
where  $V_0, V_0'', V_0''', \ldots$ are, respectively, the value and higher derivatives of the potential $V(x)$ in the maximum, $A_k(\mathcal{K}^2)$ are polynomials in the derivatives $(\omega^2 - V)', (\omega^2 - V)'', \ldots$ whose explicit forms can be found in \cite{Konoplya:2019hlu} and $\mathcal{K}$ corresponds to $( n = 0, 1, 2, \ldots)$
\begin{equation}
    \mathcal{K} = 
    \left\{\begin{matrix}
        & +n + \frac{1}{2}, \quad  && \text{$\operatorname{Re}(\omega) > 0$;}  \\ \\
        & -n - \frac{1}{2}, \hspace{0.5cm}   && \hspace{0.2cm} \text{$\operatorname{Re}(\omega) < 0$.} 
    \end{matrix}\right.
\end{equation}

Exact analytical solutions for the quasinormal spectra of black holes are only obtained in special cases, such as when Gauss hypergeometric functions can be associated with the radial component of the wave function or when the potential $V(r)$ is of the Pöschl-Teller type \cite{Cheriyodathillathu:2022fwe,Li:2024npg}. The nontrivial form of the Dymnikova black hole metric requires the use of numerical methods for determining the quasinormal modes. Among the available methods, the WKB method stands out, being widely used in this type of analysis and adopted in the present work \cite{Matyjasek:2017psv,Bolokhov:2023ruj,Skvortsova:2023zmj}.

 By employing the Schutz, Will and Iyer numerical method \cite{Schutz:1985km,Iyer:1986np,Iyer:1986nq} a high degree of accuracy is achieved for lower-lying modes. The frequencies  associated with the lowest decaying modes,  for $\ell = 4$ and $\ell = 5$, with $r_0 = 0.3$ and $r_g = 2$, considering different values of $c$ and $w$, are presented in Tables I and II. By comparing Tables I, II and III, we observe that the real part of the frequencies in the Dymnikova black hole without quintessence is higher, while the imaginary part is smaller. This indicates that the presence of quintessence leads to more rapidly damped oscillations. The results found suggest that, within the range of parameters analyzed, the black hole remains stable under scalar perturbations, since $\operatorname{Im}\omega < 0$.

 \begin{table}[h!]
        \centering
        \caption{The low overtones quasinormal mode frequencies of a massless scalar field in the Dymnikova black hole surrounded by quintessence, with fixed parameters $\ell =4$, $w= -2/3, r_0 = 0.3$ and $r_g = 2$.}
        \begin{tabular}{c|ccccc}
            \hline\hline
            $c$ & $\omega\ (n=0)$ & $\omega\ (n=1)$ & $\omega\ (n=2)$ & $\omega\ (n=3)$ & $\omega\ (n=4)$ \\
            \hline
            0.003    & 0.84341 - 0.092933i & 0.83234 - 0.28041i & 0.81121 - 0.47258i & 0.78212 - 0.67206i & 0.74799 - 0.88048i \\
            0.004 & 0.83531 - 0.091775i & 0.82442 - 0.27690i & 0.80363 - 0.46664i & 0.77497 - 0.66354i & 0.74133 - 0.86924i \\
            0.005 & 0.82717 - 0.090613i & 0.81645 - 0.27339i & 0.79599 - 0.46068i & 0.76778 - 0.65499i & 0.73463 - 0.85796i \\
            0.006 & 0.81897 - 0.089448i & 0.80843 - 0.26986i & 0.78830 -0.45470i & 0.76053 - 0.64644i & 0.72787 - 0.84666i \\
            0.007 & 0.81071 - 0.088280i & 0.80035 - 0.26632i & 0.78056 - 0.44871i & 0.75324 - 0.63785i & 0.72107 - 0.83533i \\
            0.008 & 0.80241 - 0.087108i & 0.79223 - 0.26278i & 0.77277 - 0.44270i & 0.74589 - 0.62924i & 0.71421 - 0.82397i \\
            0.009 & 0.79405 - 0.085933i & 0.78404 - 0.25922i & 0.76492 - 0.43667i & 0.73849 - 0.62061i & 0.70730 - 0.81257i \\
            \hline\hline
    \end{tabular}
\end{table} 

\begin{table}[h!]
\centering
\caption{The low overtones quasinormal mode frequencies of a massless scalar field in the Dymnikova black hole surrounded by quintessence, with fixed parameters $l=4$, $c= 0.004, r_0 = 0.3$ and $r_g = 2$.}
\begin{tabular}{c|ccccc}
\hline\hline
$w$ & $\omega\ (n=0)$ & $\omega\ (n=1)$ & $\omega\ (n=2)$ & $\omega\ (n=3)$ & $\omega\ (n=4)$ \\
\hline
-8/9 & 0.79848 - 0.088068i & 0.78892 - 0.26549i & 0.77055 - 0.44673i & 0.74492 - 0.63402i & 0.71437 - 0.82896i \\
-7/9 & 0.82046 - 0.090097i & 0.81001 - 0.27178i & 0.79005 - 0.45782i & 0.76246 - 0.65068i & 0.72995 - 0.85193i \\
-2/3 & 0.83531 - 0.091775i & 0.82442 - 0.27690i & 0.80363 - 0.46664i & 0.77497 - 0.66354i & 0.74133 - 0.86924i \\
-5/9 & 0.84540 - 0.093103i & 0.83426 - 0.28093i & 0.81303 - 0.47347i & 0.78378 - 0.67337i & 0.74949 - 0.88227i \\
-4/9 & 0.85227 - 0.094112i & 0.84099 - 0.28398i & 0.81949 - 0.47864i & 0.78989 - 0.68076i & 0.75520 - 0.89201i \\
\hline\hline
\end{tabular}
\end{table}

\begin{table}[h!]
\centering
\caption{The low overtones quasinormal mode frequencies of a massless scalar field in the Dymnikova black hole  for fixed $l=4$, $ r_0 = 0.3$ and $r_g = 2$.}
\begin{tabular}{ccccc}
\hline\hline
 $\omega\ (n=0)$ & $\omega\ (n=1)$ & $\omega\ (n=2)$ & $\omega\ (n=3)$ & $\omega\ (n=4)$ \\
\hline
 0.86742 - 0.096392i & 0.85581 - 0.29088i & 0.83369 - 0.49032i & 0.80328 - 0.69749i & 0.76771 - 0.91406i \\
\hline\hline
\end{tabular}
\end{table}

\begin{table}[h!]
\centering
\caption{The low overtones quasinormal mode frequencies of a massless scalar field in the Dymnikova black hole surrounded by quintessence, with fixed parameters $\ell =5$, $w= -2/3, r_0 = 0.3$ and $r_g = 2$.}
\begin{tabular}{c|ccccc}
\hline\hline
$c$ & $\omega\ (n=0)$ & $\omega\ (n=1)$ & $\omega\ (n=2)$ & $\omega\ (n=3)$ & $\omega\ (n=4)$ \\
\hline
0.003    & 1.03044 - 0.092873i & 1.02130 - 0.27970i & 1.00359 - 0.46972i & 0.97845 - 0.66484i & 0.94758 - 0.86658i \\
0.004 & 1.02059 - 0.091712i & 1.01160 - 0.27620i & 0.99418 - 0.46381i & 0.96942 - 0.65643i & 0.93900 - 0.85556i \\
0.005 &1.01069 - 0.090549i & 1.00185 - 0.27269i & 0.98470 - 0.45789i & 0.96033 - 0.64801i & 0.93037 - 0.84451i \\
0.006 & 1.00072 - 0.089382i & 0.992023 - 0.26917i & 0.97515 - 0.45196i & 0.95118 - 0.63957i & 0.92168 - 0.83344i \\
0.007 & 0.99069 - 0.088212i & 0.98214 - 0.26564i & 0.96555 - 0.44601i & 0.94196 - 0.63110i & 0.91292 - 0.82233i \\
0.008 & 0.98058 - 0.087039i & 0.972182 - 0.26210i & 0.95588 - 0.44004i & 0.93268 - 0.62261i & 0.90410 - 0.81119i \\
0.009 & 0.97042 - 0.085863i & 0.96216 - 0.25855i & 0.94614 - 0.43405i & 0.92333 - 0.61409i & 0.89521 - 0.80003i \\
\hline\hline
\end{tabular}
\end{table}

\begin{table}[h!]
\centering
\caption{The low overtones quasinormal mode frequencies of a massless scalar field in the Dymnikova black hole surrounded by quintessence, with fixed parameters $l=5$, $c= 0.004, r_0 = 0.3$ and $r_g = 2$.}
\begin{tabular}{c|ccccc}
\hline\hline
$w$ & $\omega\ (n=0)$ & $\omega\ (n=1)$ & $\omega\ (n=2)$ & $\omega\ (n=3)$ & $\omega\ (n=4)$ \\
\hline
-8/9 & 0.79848 - 0.088068i & 0.78892 - 0.26549i & 0.77055 - 0.44673i & 0.74492 - 0.63402i & 0.71437 - 0.82896i \\
-7/9 & 1.00260 - 0.090018i & 0.99398 - 0.27106i & 0.97723 - 0.45506i & 0.95341 - 0.64383i & 0.92406 - 0.83878i \\
-2/3 & 1.02059 - 0.091712i & 1.01160 - 0.27620i & 0.99418 - 0.46381i & 0.96942 - 0.65644i & 0.93900 - 0.85556i \\
-5/9 & 1.03282 - 0.093047i & 1.02364 - 0.28023i & 1.00584 - 0.47062i & 0.98057 - 0.66614i & 0.94955 - 0.86832i \\
-4/9 & 1.04116 - 0.094058i & 1.03186 - 0.28328i & 1.01384 - 0.47576i & 0.98826 - 0.67344i & 0.95686 - 0.87789i \\
\hline\hline
\end{tabular}
\end{table}

\begin{table}[h!]
\centering
\caption{The low overtones quasinormal mode frequencies of a massless scalar field in the Dymnikova black hole  for fixed $l=5$, $ r_0 = 0.3$ and $r_g = 2$.}
\begin{tabular}{ccccc}
\hline\hline
 $\omega\ (n=0)$ & $\omega\ (n=1)$ & $\omega\ (n=2)$ & $\omega\ (n=3)$ & $\omega\ (n=4)$ \\
\hline
 1.05961 - 0.096337i & 1.05004 - 0.29015i & 1.03150 - 0.48734i & 1.00520 - 0.68993i & 0.97296 - 0.89949i \\
\hline\hline
\end{tabular}
\end{table}

\section{Conclusion}

In this work, we investigate the properties of a Dymnikova BH immersed in a quintessential field, characterized by the state parameter $\omega$ and a normalization constant $c$. We explore the thermodynamic behavior, null geodesics, scalar quasinormal modes and shadow profiles for this model.

Regarding the thermodynamic analysis, we could see that the presence of quintessence introduces changes in the Hawking temperature and heat capacity, where phase transitions become dependent on the model parameters and reduces to the standard Dymnikova case when $c = 0$. The Hawking temperature first increases as the horizon radius grows, peaks at a certain point, and then gradually decreases as the horizon radius continues to expand. The peak temperature value increases as the $\omega$ parameter increases. Also, it was observed that the peak Hawking temperature decreases as $c$ increases, with $r_g, r_0$ and $w$ fixed. We have also shown that heat capacity, and consequently the black hole thermodynamic stability, is strongly dependent on quintessence parameters $(\omega, c)$. 

The analysis of geodesics showed how quintessence affects light deflection and the formation of shadows. By means of a consistent numerical analysis we could notice that, for any given value of $c$, as the $\omega$ parameter gets smaller the black hole tends to enhance the confinement of the particles. The opposite behavior occurs if we consider a fixed $\omega$. In this case, the increasing values of the parameter $c$ intensify the confinement of the particles.

Finally, in order to investigate the stability of the Dymnikova black hole surrounded by quintessence under scalar perturbations, we have computed the quasinormal modes by employing the WKB method. In turn, the quasinormal modes revealed that the addition of quintessence increases the stability of the solution under scalar perturbations, since it leads to more rapidly damped oscillations. 

Through this theoretical investigation, we hope to offer useful insights that bring us closer to a deeper and clearer understanding of regular black holes, specially the Dymnikova one, under the influence of a quintessence field.

\acknowledgments  
\hspace{0.5cm} The authors thank the Coordena\c{c}\~{a}o de Aperfei\c{c}oamento de Pessoal de N\'{i}vel Superior (CAPES). JF would like to thank Alexandra Elbakyan and Sci-Hub, for removing all barriers in the way of science and the Funda\c{c}\~{a}o Cearense de Apoio ao Desenvolvimento Cient\'{i}fico e Tecnol\'{o}gico (FUNCAP) under the grant PRONEM PNE0112- 00085.01.00/16 and the Conselho Nacional de Desenvolvimento Científico e Tecnol\'{o}gico (CNPq) under the grant 304485/2023-3.

 \end{document}